\documentclass[twocolumn, switch]{article} 

\usepackage{preprint}
\usepackage{amsmath, amsthm, amssymb, amsfonts}
\usepackage[numbers,square]{natbib}
\usepackage[utf8]{inputenc}	
\usepackage[T1]{fontenc}	
\usepackage{xcolor}		
\usepackage[colorlinks = true,
            linkcolor = purple,
            urlcolor  = blue,
            citecolor = cyan,
            anchorcolor = black]{hyperref}	
\usepackage{booktabs} 		
\usepackage{nicefrac}		
\usepackage{microtype}		
\usepackage{lineno}		
\usepackage{float}			

\usepackage{lipsum}		
\usepackage{url}

\usepackage[binary-units=true]{siunitx}
\usepackage[left]{eurosym}
\def\code#1{\texttt{#1}}
\usepackage{newfloat}
\DeclareFloatingEnvironment[name={Supplementary Figure}]{suppfigure}
\usepackage{sidecap}
\sidecaptionvpos{figure}{c}

\usepackage{titlesec}
\titlespacing\section{0pt}{12pt plus 3pt minus 3pt}{1pt plus 1pt minus 1pt}
\titlespacing\subsection{0pt}{10pt plus 3pt minus 3pt}{1pt plus 1pt minus 1pt}
\titlespacing\subsubsection{0pt}{8pt plus 3pt minus 3pt}{1pt plus 1pt minus 1pt}

\usepackage{tikz,xcolor,hyperref}

\definecolor{lime}{HTML}{A6CE39}
\DeclareRobustCommand{\orcidicon}{
	\begin{tikzpicture}
	\draw[lime, fill=lime] (0,0)
	circle [radius=0.16]
	node[white] {{\fontfamily{qag}\selectfont \tiny ID}};
	\draw[white, fill=white] (-0.0625,0.095)
	circle [radius=0.007];
	\end{tikzpicture}
	\hspace{-2mm}
}
\foreach \x in {A, ..., Z}{\expandafter\xdef\csname orcid\x\endcsname{\noexpand\href{https://orcid.org/\csname orcidauthor\x\endcsname}
			{\noexpand\orcidicon}}
}

\title{SonoTraceLab - A Raytracing-Based Acoustic Modelling System for Simulating Echolocation Behavior of Bats}

\usepackage{authblk}

\author[1,2]{Wouter Jansen\orcidA{}}
\author[1,2,3]{Jan Steckel\orcidB{}}

\affil[1]{Cosys-Lab, Faculty of Applied Engineering, University of Antwerp, Antwerp, Belgium}
\affil[2]{Flanders Make Strategic Research Centre, Lommel, Belgium}
\affil[3]{Corresponding Author: jan.steckel@uantwerpen.be}

\begin{document}

\twocolumn[ 
  \begin{@twocolumnfalse} 
\maketitle


\begin{abstract}
Echolocation is the prime sensing modality for many species of bats, who show the intricate ability to perform a plethora of tasks in complex and unstructured environments. Understanding this exceptional feat of sensorimotor interaction is a key aspect into building more robust and performant man-made sonar sensors. In order to better understand the underlying perception mechanisms it is important to get a good insight into the nature of the reflected signals that the bat perceives. While ensonification experiments are in important way to better understand the nature of these signals, they are as time-consuming to perform as they are informative. In this paper we present SonoTraceLab, an open-source software package for simulating both technical as well as biological sonar systems in complex scenes. Using simulation approaches can drastically increase insights into the nature of biological echolocation systems, while reducing the time- and material complexity of performing them.

\end{abstract}
\vspace{0.35cm}

  \end{@twocolumnfalse} 
] 


\section{Introduction}
Bats exhibit a wide range of behaviors that are of interest to scientists who want to understand the underlying mechanisms of perceiving the world using peculiar sensing modalities \citep{schnitzler2003spatial}. Indeed, bats rely mainly on echolocation to perceive their world, which poses an entirely different set of challenges to these animals compared to animals relying on vision to navigate their surroundings \citep{Geipel2019, prat2020decision, stidsholt2021hunting, mcgowan2020different}. To better understand this echolocation behavior, experiments have been performed to understand better the environments bats encounter, and how these environments manifest themselves in the sensory domain of bats. For example, nectar-feeding bats have been shown to be attracted to specialized leaves formed by neo-tropical pitcher plants \citep{simon2006size, simon2011floral, simon2021acoustic, schoner2015bats}. Similarly, the neo-tropical bat Micronycteris \textit{microtus} can hunt insects in dense vegetation \citep{Geipel2019, geipel2020predation, geipel2013perception}, and smooth surfaces have been shown to attract bats, as the echolocation signature of such surfaces resembles bodies of water \citep{stilz2017glass, greif2010innate}. All these aforementioned studies have in common that the animals clearly react to specific properties of the scene being ensonified, and therefore, to understand the behavior, the nature and intricate details of the reflected signals should be well understood.
\\\\
In order to better understand the echolocation behavior of bats, a plethora of ensonification experiments have been performed. Indeed, for static scenes such as the foraging behavior of bats close to pitcher plants \citep{schonerBatsAreAcoustically2015} or bats hunting motionless prey \citep{geipel2013perception, moto:c:irua:161459_geip_bats}, and the hunting behavior of trawling bats who hunt over water surfaces \citep{ubernickelSensoryChallengesTrawling2016}, ensonification experiments have revealed insights into the underlying mechanisms that allow bats to solve these complex tasks successfully. Furthermore, in dynamic scenes, such as the recognition of fluttering prey items, ensonification experiments have yielded profound insights into potential signal processing schemes that underpin the hunting strategies \citep{fontaineCompressiveSensingStrategy2011, mossAcousticInformationAvailable1994}. These successes strongly advocate for the performance of these ensonification experiments. However, ensonification experiments come with significant drawbacks. Firstly, these experiments are complex to perform from both a mechanical and an acoustic point of view. Acoustically, it is important to accurately reproduce the external echolocation peripherals of the bats to yield relevant information about the interaction of the reflector filtering and the external acoustic filtering (called the Head-Related Transfer Function, HRTF) of the bat's emission and reception organs. Approaches to solving this have been found in implementing artificial, 3D printed pinnae \citep{caspersDesignDynamicBiomimetic2018, schillebeeckxBiomimeticSonarBinaural2011a, sutliveBiomimeticSoftRobotic2020}, or using arrays of microphones and emitters to accurately replicate the directivity patterns of the bat \citep{moto:c:irua:100095_stec_nove, moto:c:irua:92037_stec_biom}. Furthermore, some bats utilize extremely high bandwidths in their echolocation calls \citep{siemersEcholocationSignalsReflect2004}, up to and beyond 180\si{\kilo\hertz}, which poses significant issues in practical implementations of both emitters and receivers.
\\\\
Mechanically, the full 6-DOF (Degrees Of Freedom) of the bat in relationship to the ensonified object needs to be controlled accurately. Typically, this is done using some kind of robotic setup using either a custom-made rig \cite{moto:c:irua:184867_simo_acou} or standard robotic arms \cite{nguyenSensorimotorBehaviorInformational2021}. However, these kind of setups introduce all kinds of additional complexity, from the ill-posedness of inverse kinematics \cite{arimotoNaturalResolutionIllposedness2005} to unwanted additional reflections from the environment occuring and tainting the experimental data. 
\\\\

\begin{figure*}
    \centering
    \includegraphics[width=0.8\linewidth]{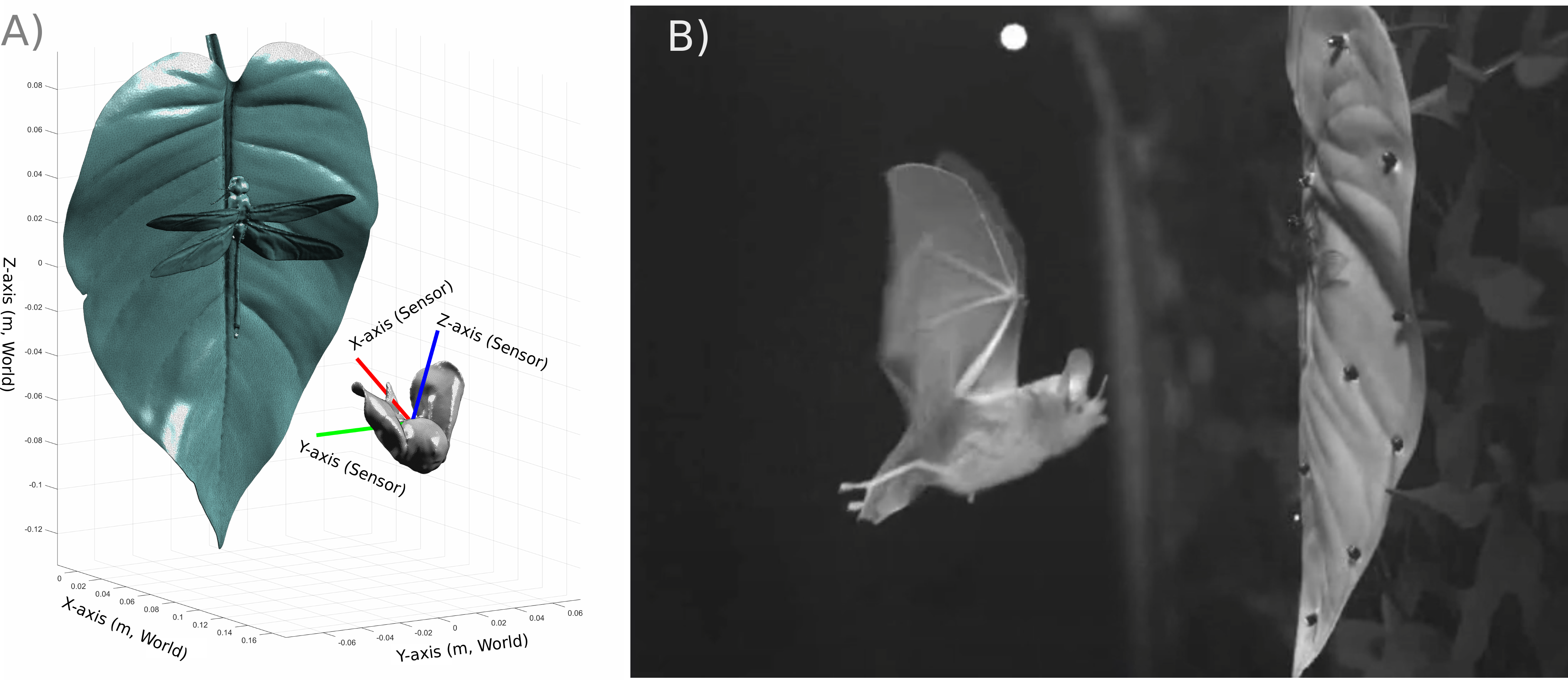}
    \caption{A typical setup for an echolocation experiment. Panel A shows the Global X, Y, and Z axes, the local X, Y, and Z axes of the echolocation sensor (red, green, and blue arrows respectively), the 3D model of the echolocation sensor (in this case a model of the head of the bat Micronycteris \textit{microtus} and the object to be ensonified, which in this case is a generic leaf with a model of a dragonfly. Panel B shows a still frame from a high-speed video showing a Micronycteris \textit{microtus} performing the same task \citep{cosys-labMicronycterisMicrotisCapturing2018}).}
    \label{fig:generalsetup}
\end{figure*}

To overcome the previously detailed limitations, one can resort to simulation approaches. Indeed, acoustic wave theory is a well-understood field \citep{pierceAcousticsIntroductionIts2019}, and many successful simulation paradigms exist. Most notably, Finite Element Methods (FEM) \citep{everstineFiniteElementFormulatons1997, ihlenburgFiniteElementAnalysis1998, thompsonReviewFiniteelementMethods2006}, Boundary Element Methods (BEM) \citep{kirkupBoundaryElementMethod2007, kirkupBoundaryElementMethod2019} and Ray acoustics \citep{onakaDesign3dimensionalSound2009, roberHRTFSimulationsAcoustic2006, roberRayAcousticsUsing2007} are typical approaches towards solving the Helmholtz equations resulting from the acoustic simulation setup. Furthermore, commercial simulation packages such as Comsol \citep{comsolcomsolCOMSOLMultiphysicsSoftware} or Siemens Simcenter Acoustics \citep{siemensSiemensNXAcoustics} allow solving acoustical problems with one or multiple of the aforementioned simulation approaches. These simulation packages are often expensive to license and complex because of their broad applicability to a wide array of problem types. Alternatively, open-source simulation engines exist, such as k-Wave \citep{treebyKWaveMATLABToolbox2010} or FIELD-II \citep{field-iiFieldIIUltrasound}. While these solutions are free to use, they also are quite broad in their application domain and, due to their computational approach towards solving the Helmholtz equation, are limited to relatively small problems in terms of the size of the scene compared to the wavelengths that are to be simulated.
\\\\\
In order to overcome the limitations of these existing solutions, we set out to develop SonoTraceLab. This acoustic simulation engine (1) is open-source, (2) allows simple imports of arbitrary geometry without going through complex meshing procedures, (3) is tailored towards solving problems typically occurring in the investigation of active ultrasound sensing (both for biological as well as robotic ultrasound systems) and (4) allows the simulation of medium-sized environments (mesh sizes up to a million triangles). SonoTraceLab also allows for including arbitrary HRTF patterns in a post-processing step, a feature important for analyzing echolocation behavior, which is not feasible in currently available simulators. In what follows, we will first expand on the simulation paradigm powering SonoTraceLab, detailing how the simulation engine functions and its inputs and outputs. Next, we will demonstrate some typical scenarios encountered in bat echolocation, which will be validated using real-world measurements. Finally, we will demonstrate how our open-source implementation can be used, and then draw some conclusions and discuss the limitations of our work. 

\section{simulation approach}

A typical simulation setup can be seen in figure \ref{fig:generalsetup}. An echolocating agent (in this case, the bat Micronycteris \textit{microtus} approaches a complex target, which in this case is a leaf with a dragonfly. The world coordinate system $[X_w, Y_w, Z_w]$ is shown, to which the overall simulation is tied. For the remainder of this paper, we will omit the subscript $W$ for brevity. The coordinate system of the sensor, $[X_s, Y_s, Z_s]$, is also shown, which is the overall coordinate system to which the sensor-specific components and behaviors are tied. This coordinate system can be placed into the environment and rotated around its axes using a 4x4 transformation matrix $T_{s,w}$ containing the rotation matrices and translation matrix. Panel B shows a still frame from a high-speed video of the same bat performing an approach to the leaf in a similar scenario. The goal of the simulation is to simulate the signals at the tympanum of the bat's ear during this approach phase, as these can then give insights into how the echolocation features the bat receives and might make use of to solve this hunting task.

\section{implementation details}
\label{sec:implementation}
\begin{figure*}
    \centering
    \includegraphics[width=\linewidth]{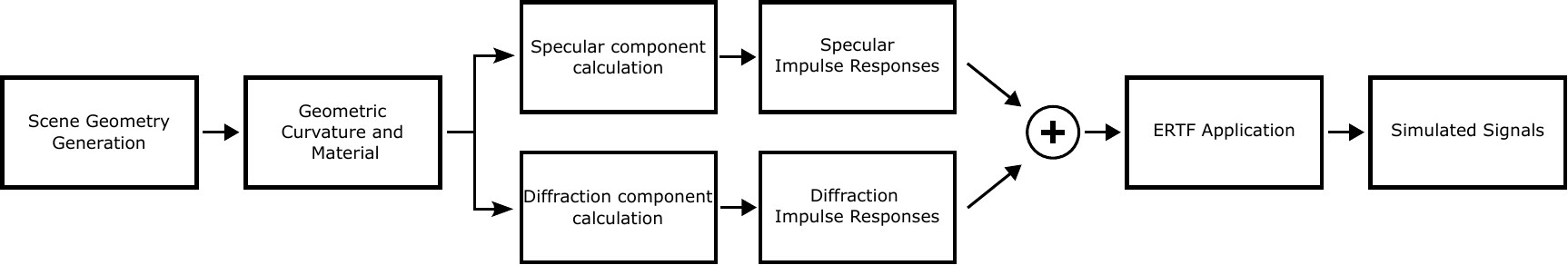}
    \caption{Block diagram illustrating the sequential stages of the simulation implementation for synthesizing acoustic signals.} 
    \label{fig:sim_schematic}
\end{figure*}

The overall flowchart of the simulation can be found in figure \ref{fig:sim_schematic}. The simulation starts by importing scene geometry, encoded in the STL format \citep{adobeSTLFilesExplained}. Once the scene geometry is loaded, the mesh is analyzed, and minor reparations on the mesh are performed to make it watertight and ensure all the normals are pointing toward the outside of the mesh. Next, the local curvature of the mesh is calculated, which is used later in the simulation process to account for diffraction effects. Then, the simulation graph bifurcates into two parts: a part for raytracing and a part for diffraction. In the raytracing part, the solution for the Helmholtz equation is found using a ray acoustics approach, which assumes large Helmholtz numbers to be correct, from which the spatial impulse response is calculated. However, as in scenes similar to the one described in figure \ref{fig:generalsetup}, the assumption of high Helmholtz numbers is not entirely correct, and therefore, diffraction effects cannot be neglected. To avoid solving the complete Helmholtz equation using FEM or BEM approaches, we resort to a Monte Carlo technique for approximating the diffraction echoes arising from the scene, using information about the local curvature of the mesh to locate diffraction echo candidates. This process calculates a spatial impulse response for the diffraction component.
\\\\
Subsequently, as we only consider linear acoustic phenomena (a valid assumption for the sound pressure levels we consider in typical echolocation scenarios), the complete spatial impulse response can be obtained by summing the specular and diffraction spatial impulse responses. Then, as the echolocating agent under consideration does not have an isotropic emission and reception pattern, we apply a spatial ERTF (Echolocation Related Transfer Function, which is the product of the emission pattern and the HRTF) to the spatial impulse response using an array-based approximation approach, first described in the echolocation context in \cite{moto:c:irua:100095_stec_nove}. Finally, we convolve the ERTF-filtered spatial impulse responses with the echolocation call of the agent, and we obtain the signals at the left and right tympanum of the bat. In what follows, we will provide more details on each step in the computational graph.

\subsection{Model preparation}
\begin{figure}
    \centering
    \includegraphics[width=\linewidth]{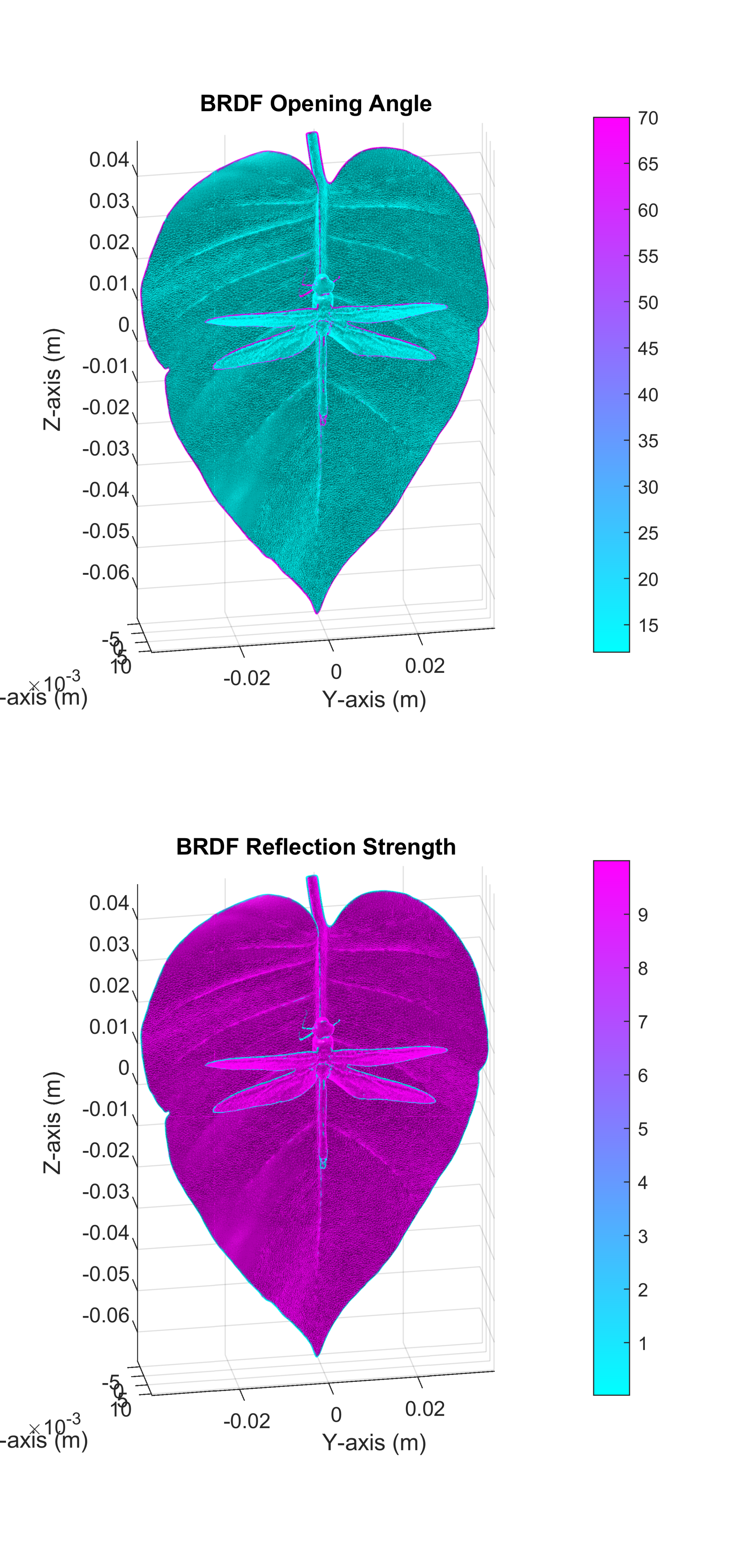}
    \caption{Overview of the acoustic BRDF (Bidirectional Reflectance Distribution Function) applied to the loaded mesh using local curvature to modulate the parameters $\alpha$ and $k$. The top panel shows the opening angle of the BRDF (i.e., how specular or omnidirectional a reflection on the local vertex is) and the reflectance strength (i.e., how strong a reflection is from this vertex). It should be noted that the opening angle towards areas with high curvature shows a significantly wider opening angle, which is desired because diffraction echoes arise at locations with high curvature and manifest themselves as almost omnidirectional reflections. However, these diffraction echoes should be significantly weaker than the specular reflected echoes, which are reflected in the BRDF reflection strength plot.}
    \label{fig:curvature}
\end{figure}

As stated before, the model mesh is stored in an STL file, which the simulation engine loads into a vertex matrix $V_m$ and a face matrix $F_m$. Next, we perform several mesh cleaning functions to remove non-manifold faces, duplicate faces, and duplicate vertices. For this, we use the Matlab Lidar Toolbox \citep{matlabRemoveSurfaceMesh}, which, if wanted, other functions could replace to achieve the same goal. Next, we calculate the surface normals $N_V$ for each vertex, which are then combined into the surface normals for each face $N_F$ through averaging. This information is then used to calculate the local curvature tensor $C$ using the approach described in \cite{rusinkiewiczEstimatingCurvaturesTheir2004}. From this tensor, we calculate the magnitude of the curvature $C_m$ per face. From this magnitude, we calculate two material properties of the scene mesh: the opening angle $\alpha$ of the acoustic BRDF (Bidirectional Reflectance Distribution Function) and the magnitude $k$ (i.e., reflection strength) of the BRDF. These values (i.e., $\alpha$ and $k$) are frequency-dependent and must be set to appropriate values for the simulation to yield realistic results. It should be noted that one could perform experiments to obtain measured values for these BRDF functions, using approaches described in \cite{binekApplicationFittedMeasurements2019, duranyAnalyticalComputationAcoustic2015, matusikEfficientIsotropicBRDF2003}, which, however, falls outside of the scope of this paper. The parameter $\alpha(v,f)$ encodes how wide of a directivity pattern the reflection from that vertex for that frequency is (i.e., how specular or omnidirectional that reflection is), and the parameter $k(v,f)$ encodes how strong the reflection of vertex $v$ is. The resulting BRDF functions can be seen in figure \ref{fig:curvature}. Panel A shows the opening angle $\alpha$ of the BRDF, and panel B shows the reflection strength $k$.
\\\\
At this stage, SonoTraceLab supports a single emitter and multiple receivers, as this is a simulation setup which is applicable for many application scenarios, both in engineered sonar sensors \citep{moto:c:irua:165188_kers_a, moto:c:irua:166446_kers_3d, moto:c:irua:168766_vere_urti, moto:c:irua:186389_bale_sens} as in biologically relevant echolocation setups, as both bats and dolphins can be approximated by a point-source with a known directivity \citep{thomasEcholocationBatsDolphins2004}. The emitter and the receiver arrays are defined in the local sensor coordinate system  $[X_s, Y_s, Z_s]$, as shown in figure \ref{fig:generalsetup}. Figure \ref{fig:raytracing}, panel A) shows the coordinate system in more detail. In this case, the emitter is located at the origin of the sensor coordinate system, and two groups of receiver arrays are made, resembling a left and a right ear. Another layout can be seen in figure \ref{fig:ERTFComparison}, where a 2D array is used for both ears, and the emitter is located below the origin. The whole sensor system can be located using a translation $T$ and rotation $R$, which is combined into the transformation matrix $T_{s,w}$, and which is parameterized by the position of the sensor in world coordinates and the rotations $[\gamma_x, \gamma_y, \gamma_z ]$ around the $X$, $Y$ and $Z$ axes respectively.

\subsection{Impulse Response Calculation using Specular Reflections}
\begin{figure*}
    \centering
    \includegraphics[width=\linewidth]{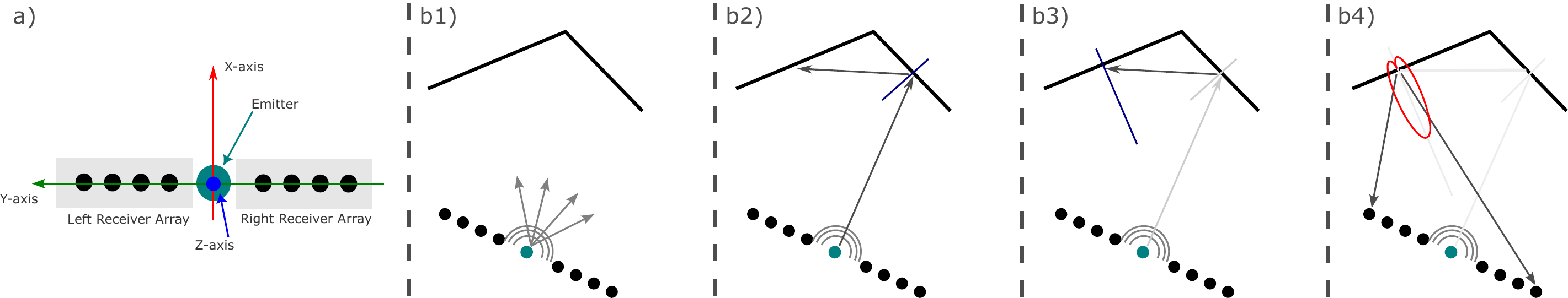}
    \caption{Panel a) shows the used coordinate system with two receiver arrays for the left and right ear, a single source, and the coordinate system attached to the sensor. Panel b shows an overview of the raytracing-based partial solution to the Helmholtz equation. In panel b1, a set of rays are cast into all directions of the frontal hemisphere. Panel b2 shows the intersection of one ray with the scene geometry, which is reflected around the surface normal. The ray further propagates (b3) and reflects from the scene geometry a final time (b4). The BRDF (red ellipse) is sampled by each microphone, depicted by two black lines intersecting the ellipse. } 
    \label{fig:raytracing}
\end{figure*}

In this section, we will discuss the raytracing process used to solve the specular part of the Helmholtz equation. The goal is to calculate the impulse response $h_i(t)$ for the i-th microphone from source $s$, propagated throughout the environment. To achieve this, we break the problem into a raytracing problem, where $N$ rays are emitted into the environment (typical values for $N$ are around 300.000 in our experiments). For each of the $N$ rays, we calculate the propagation path throughout the environment using a specular reflection model for each model face. The overall process can be seen in figure \ref{fig:raytracing}, panels B1 through B4. It should be noted that while the figure shows a 2D view of the problem, the simulation works in 3D. Intersections of cast and reflected rays are calculated using a custom-made GPU implementation of the Moller-Trumbore algorithm \citep{shumskiyGpuRayTracing2013}, implemented in CUDA (for achieving faster computation) and linked to Matlab through CUDA-MEX. In the final panel (B4) of figure \ref{fig:raytracing}, two rays from the last reflection point are shown, one to each microphone. Each of these rays samples another value of the red BRDF function, yielding different reflection strengths and path lengths due to the geometric layout of the problem. The raytracing process yields a transfer function per ray, called $H(f,n,i)$, which encodes the complex transfer function between the source and the $i$-th microphone for the $n$-th ray for each frequency $f$. This transfer function is calculated as follows:
$$
H(f,n,i) = H_m(f,n,i) \cdot e^{j k_w r_{i,n} } \cdot \frac{1}{r_{i,n}^2}
$$
where $H_m$ represents the magnitude of the reflection and is calculated from the BRDF functions of the reflections during the raytracing process, and the term $e^{j k_w r_{i,n} }$ represents the delay of the signal due to path propagation. The term $\frac{1}{r_{i,n}^2}$ represents the path loss of the signal due to spherical spreading. This complex transfer function $H(f,n,i)$ can then be transformed into the time domain using an inverse Fast Fourier Transform:
$$
h(t,n,i) = \mathcal{F}^{-1} \bigg[ H(f,n,i) \bigg]
$$
Finally, we can calculate the impulse response for the i-th microphone between source $s$, propagated through the environment, by summing the $N$ impulse responses:
$$
h_i(t) = \sum^{N}_{n=1} h(t,n,i) 
$$
Here, the impulse responses $h_i(t)$ contain the approximation to the solution of the Helmholtz equation under the assumption that the Helmholtz number is large, i.e., that specular reflections are the dominant reflection type occurring in the scene.

\subsection{Impulse Response Calculation using Diffraction}
As stated in the previous section, raytracing-based approaches are only valid for large Helmholtz numbers. However, in many situations encountered in echolocation scenarios, diffraction echoes are a significant aspect of the overall impulse response. Nevertheless, solving the full Helmholtz equation is computationally untractable for typical scene geometries. Therefore, we resort to a Monte Carlo approximation of the wave equation solution and utilize the local curvature calculated over the scene geometry to find diffraction echo candidate locations. Concretely, we randomly distribute a fixed number of points onto mesh areas with high local curvature. We use an importance sampling approach \cite{corsiniEfficientFlexibleSampling2012} based on the cumulative probability distribution to distribute these points onto the mesh. Then, we follow a similar approach to impulse response generation as in the specular reflection case. First, we calculate the ranges $r_{i,m}$ between the $i$-th microphone and the $m$-th diffraction point, added to the distance between the diffraction point and source $s$. Then, we synthesize the transfer function $G(f,m,i)$ as follows:
$$
G(f,m,i) = G_m(f,m,i) \cdot e^{j k_w r_{i,m} } \cdot \frac{1}{r_{i,m}^2}
$$
Similarly to the specular reflections, $G_m$ represents the magnitude of the reflections calculated from the BRDF angle and material property. The term $e^{j k_w r_{i,m}}$ represents the delay due to the distance between the diffraction point and the microphone, and $\frac{1}{r_{i,m}^2}$ represents the effect of path loss due to spherical spreading.  This complex transfer function $G(f,m,i)$ can then be transformed into the time domain using an inverse Fast Fourier Transform:
$$
g(t,m,i) = \mathcal{F}^{-1} \bigg[ G(f,m,i) \bigg]
$$
Finally, we can calculate the impulse response for the i-th microphone between source $s$, propagated through the environment, by summing the $N$ impulse responses:
$$
g_i(t) = \sum^{M}_{m=1} g(t,m,i) 
$$
The term $g_i(t)$ contains the partial solution to the Helmholtz equation for the diffraction components of the scene. 

\begin{figure}
    \centering
    \includegraphics[width=0.8\linewidth]{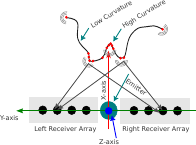}
    \caption{Overview of the approach of solving the diffraction aspects of the Helmholtz equation. We calculate the curvature using local differential geometry and sample locations with high curvature to create diffraction echoes. The generated samples then sample the local BRDF properties and provide input to the impulse response generation module.} 
    \label{fig:diffraction}
\end{figure}

\subsection{ERTF implementation}
\begin{figure}
    \centering
    \includegraphics[width=\linewidth]{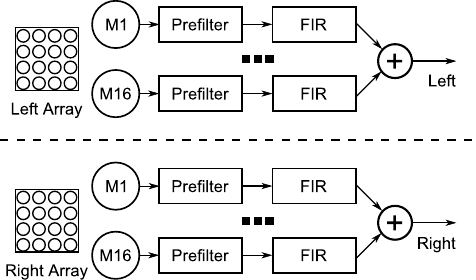}
    \caption{Overview of the array signal processing chain to implement the ERTF spatial filters. Each microphone is passed to an optional prefilter, followed by a FIR filter calculated using a least-squares model described in .} 
    \label{fig:ertffir}
\end{figure}

The sources $s$ and the receivers $m_i$ are omnidirectional in nature. However, as most real-world sensors of echolocating agents have non-ideal spatial directivity patterns, it is important that these spatial transfer functions are represented in SonoTraceLab. This is especially important for the simulations of biological echolocation systems, as the HRTF or ERTF are instrumental in the behavior of these animals \cite{demeyModellingSimultaneousEcho2010,obristWhatEarsBats1993,vanderelstWhatNoseleavesFM2010} . However, tailor-made HRTF implementations are intractable in the simulation applied in SonoTraceLab. More importantly, it might be necessary to simulate array-based sensors that adjust their directivity patterns in post-processing or bats that alter their pinna shapes during echolocation. Therefore, we opted to apply an array-based approach to implementing ERTFs, similar to the approach we presented in \cite{moto:c:irua:100095_stec_nove}. In that work, we presented an ERTF fitting process to an arbitrary array of microphones using a least-squares approach and a far-field array response model. The overview of the signal processing process is shown in figure \ref{fig:ertffir}. For a given ERTF $E(f,\psi)$, parameterized by frequency $f$ and direction $\psi$, a set of FIR filters $h_e(t,i)$ can be synthesized that, once applied to each microphone and summed, result in a single signal with the desired spatial directivity pattern. For example, for an arbitrary set of signals $s_i(t)$, the computation to apply an ERTF is as follows:
$$
s_f(t) = \sum^{I}_{i=1} h_e(t,i) * s_i(t) 
$$

\subsection{Acoustic Signal Generation}
At this point, we have all the necessary parts to implement the final step of the simulation, namely the generation of the signals at the tympanum of the bat. First, we need to generate the complete spatial impulse response of the scene by combining the partial solutions obtained from the raytracing module and the diffraction module:
$$
h_t(t,i) = a_r \cdot h_i(t) + a_d \cdot g_i(t)
$$
with $a_r$ and $a_d$ gain factors for the ray-acoustics and the diffraction parts of the signals. Then, we filter these signals using the ERTF FIR filters to obtain a spatially filtered spatial impulse response, both for the left ear ($h_l(t,i)$) and the right ear ($h_r(t,i)$):
\begin{align*}
h_l(t) &= \sum^{I}_{i=1} h_{e,l}(t,i) * h_t(t,i) \\
h_r(t) &= \sum^{I}_{i=1} h_{e,r}(t,i) * h_t(t,i)
\end{align*}
Now, with assuming a signal $s_e(t)$ emitted by the source $s$, we can calculate the signals arriving at the tympanum of the bat as follows:
\begin{align*}
s_l(t) &= h_l(t) * s_e(t) \\
s_r(t) &= h_r(t) * s_e(t)
\end{align*}
With the generation of these signals, we have reached the end of the mathematical description of the simulation process, and we want to congratulate the reader on making it to this point in this paper. What will follow is a set of validation experiments showing the functionality of SonoTraceLab.

\section{validation}
So far in this paper, we have shown how SonoTraceLab is implemented and how an approximation of the solution to the acoustic Helmholtz equation can be found in a computationally efficient manner. In this section, we will delve deeper into the practical applications of SonoTraceLab and present several examples of how SonoTraceLab can be used in biologically relevant echolocation scenes. Concretely, we will demonstrate three scenarios. In the first scenario, we will show how the ERTF implementation can yield sensor arrays with a spatial directivity function similar to the ERTF of a real bat. In the second scenario, we will show the spatial reflectivity function of the leaves of pitcher plants, both measured and simulated. Finally, we will show an example of the hunting behavior of Micronycteris \textit{microtus}, approaching a leaf with a dragonfly.

\subsection{Spatial filtering through ERTF fitting}
\begin{figure*}
    \centering
    \includegraphics[width=\linewidth]{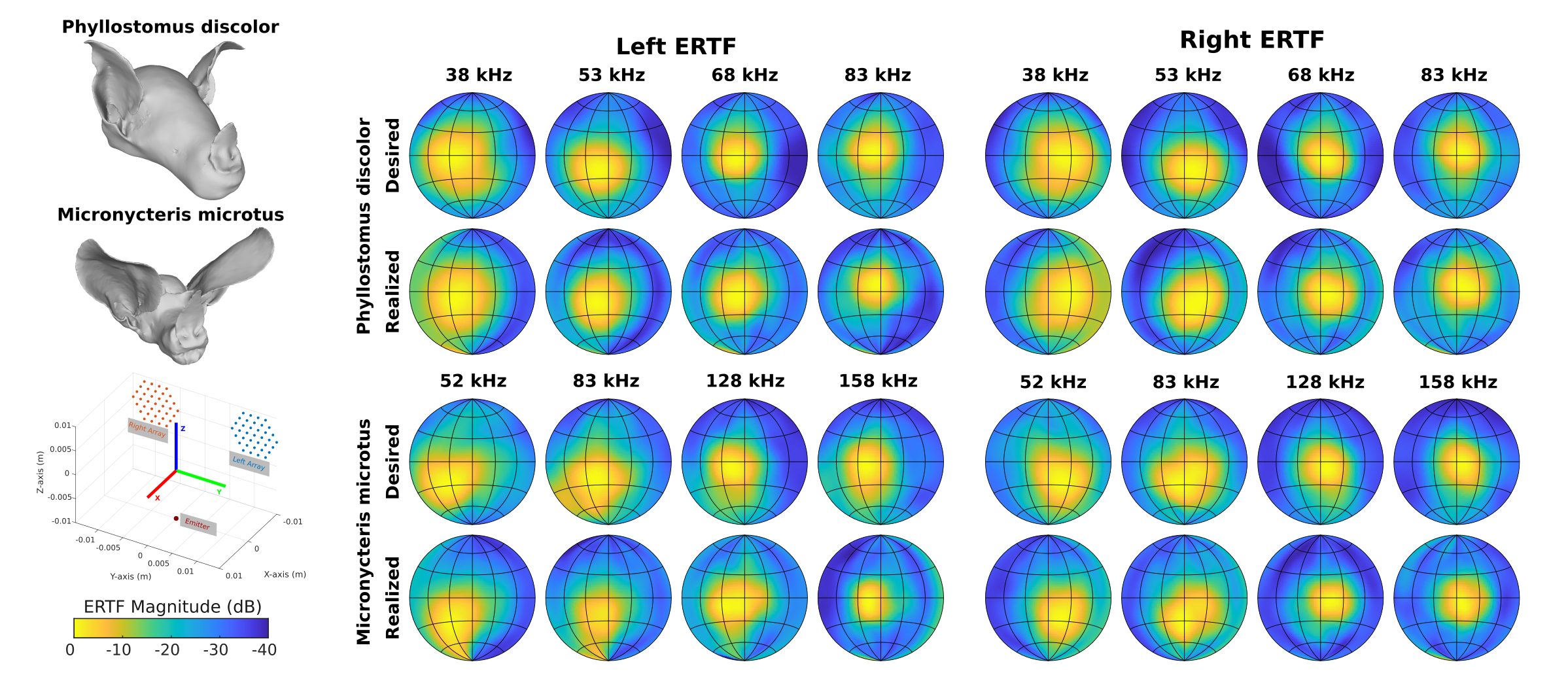}
    \caption{Overview of the validation experiment on fitting arbitrary ERTFs. This figure shows the results of fitting two ERTFs of two bats (Phyllostomus \textit{discolor} and Micronycteris \textit{microtus} on the microphone array shown in the bottom left of the figure. The top left shows the 3D models of the heads of these bats (not to scale). We show both the desired ERTFs (simulated using a Boundary Element Method), and the realized ones coming from the SonoTraceLab simulation.} 
    \label{fig:ERTFComparison}
\end{figure*}

As discussed in the previous sections, we have implemented an array-based spatial filtering method that allows the implementation of virtually any ERTF in post-processing. To validate this approach, we simulated a scene with a single sphere of 5\si{\centi\meter} in front of a sensor as shown in figure \ref{fig:ERTFComparison}, bottom left corner. The sphere was placed on 1000 points, distributed uniformly using an equal area sphere partitioning algorithm on the \citep{leopardiPartitionUnitSphere2006} frontal hemisphere with a radius of 1\si{\meter}. Then, we calculated the left and right impulse responses using the approach described in the previous section. Finally, we calculated the power spectral density of the reflected impulse response and plotted the so-obtained ERTFs. These results can be seen in figure \ref{fig:ERTFComparison}. The figure shows the desired (i.e., simulated using a BEM based on micro-CT scans \citep{demeySimulatedHeadRelated2008, vanderelstWhatNoseleavesFM2010}). These are shown for the bat Phyllostomus \textit{discolor} and the bat Micronycteris \textit{microtus}, for various frequency ranges, plotted using a Lambert Equal Area projection. We also plot the simulated responses from SonoTraceLab for the left and right ERTF in the same frequency ranges. It becomes apparent that these ERTFs show a significant degree of similarity, with similar degrees of correspondence as discussed in \cite{moto:c:irua:100095_stec_nove}. This experiment indicates two major functionalities of SonoTraceLab. Firstly, it indicated that the simulation approach for microphone arrays is correct and that time and phase differences of the reflected signals are correctly calculated. Secondly, it indicates that arbitrary ERTF functions can be implemented in SonoTraceLab, an important aspect for simulating the behavior of bats and dolphins.

\subsection{Reflector filtering of biologically relevant reflectors}
\begin{figure}
    \centering
    \includegraphics[width=\linewidth]{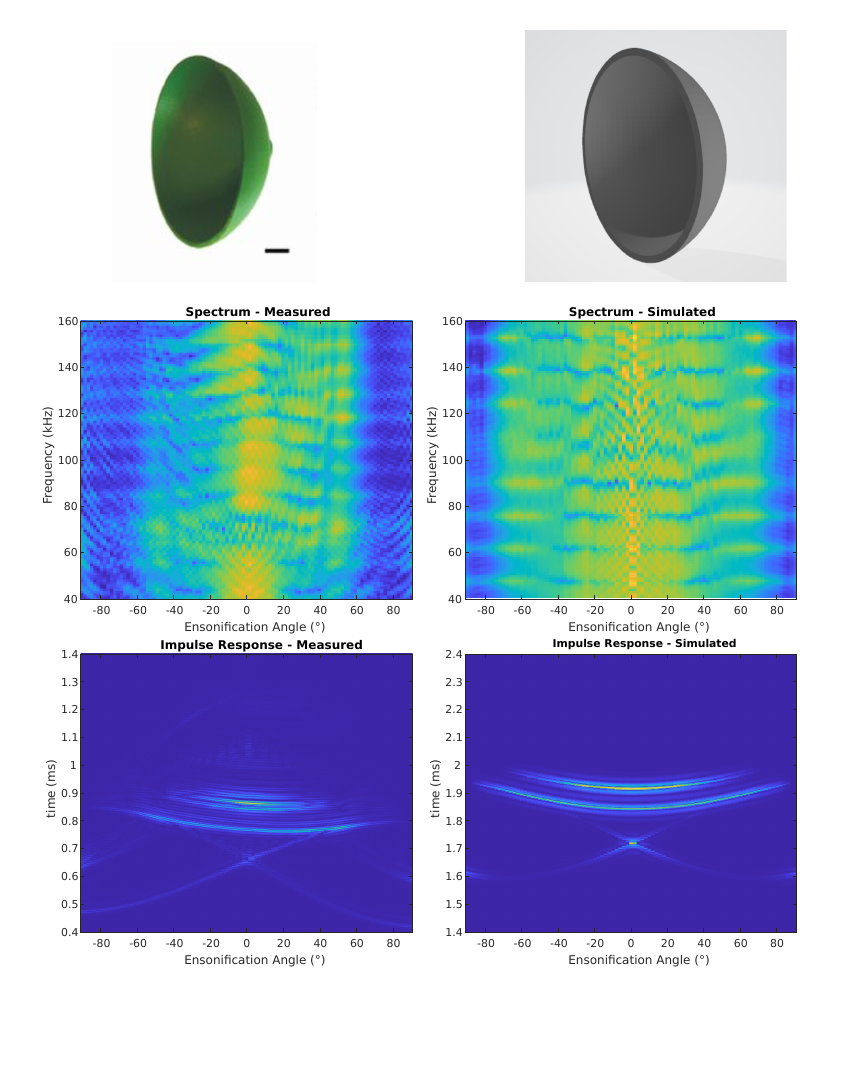}
    \caption{Spatial reflectivity patterns of a biologically inspired reflector shape, showing a distinct spatial response in the frequency domain (by means of stable notches over incidence angle). We both show the measured responses (left column) and the simulated responses using SonoTraceLab. Furthermore, we also show the time-domain impulse responses.} 
    \label{fig:biorefl}
\end{figure}

As a second validation experiment, we set out to recreate the reflectivity function of the biologically inspired reflector shapes described in our previous work \citep{moto:c:irua:165435_simo_bioi}. In that previous work, we proposed reflector shapes based on the shapes of the leaves of neo-tropical pitcher plants and used those as guiding beacons for robotic navigation. In that paper, we measured the reflectivity pattern of several 3D printed shapes, as shown in figure \ref{fig:biorefl}, left panel. The plot shows the impulse response both in the frequency domain as well as in the time domain for incidence angles ranging from -90\si{\degree} to 90\si{\degree} in steps of 1\si{\degree}. We simulated a similar setup in SonoTraceLab and calculated the same responses as in the real-world experiment. Here, we obtain very similar characteristics in both the time-domain representation of the impulse response and the frequency-domain response. 

\subsection{Target Position Discrimination using Biosonar}
\begin{figure*}
    \centering
    \includegraphics[width=0.9\linewidth]{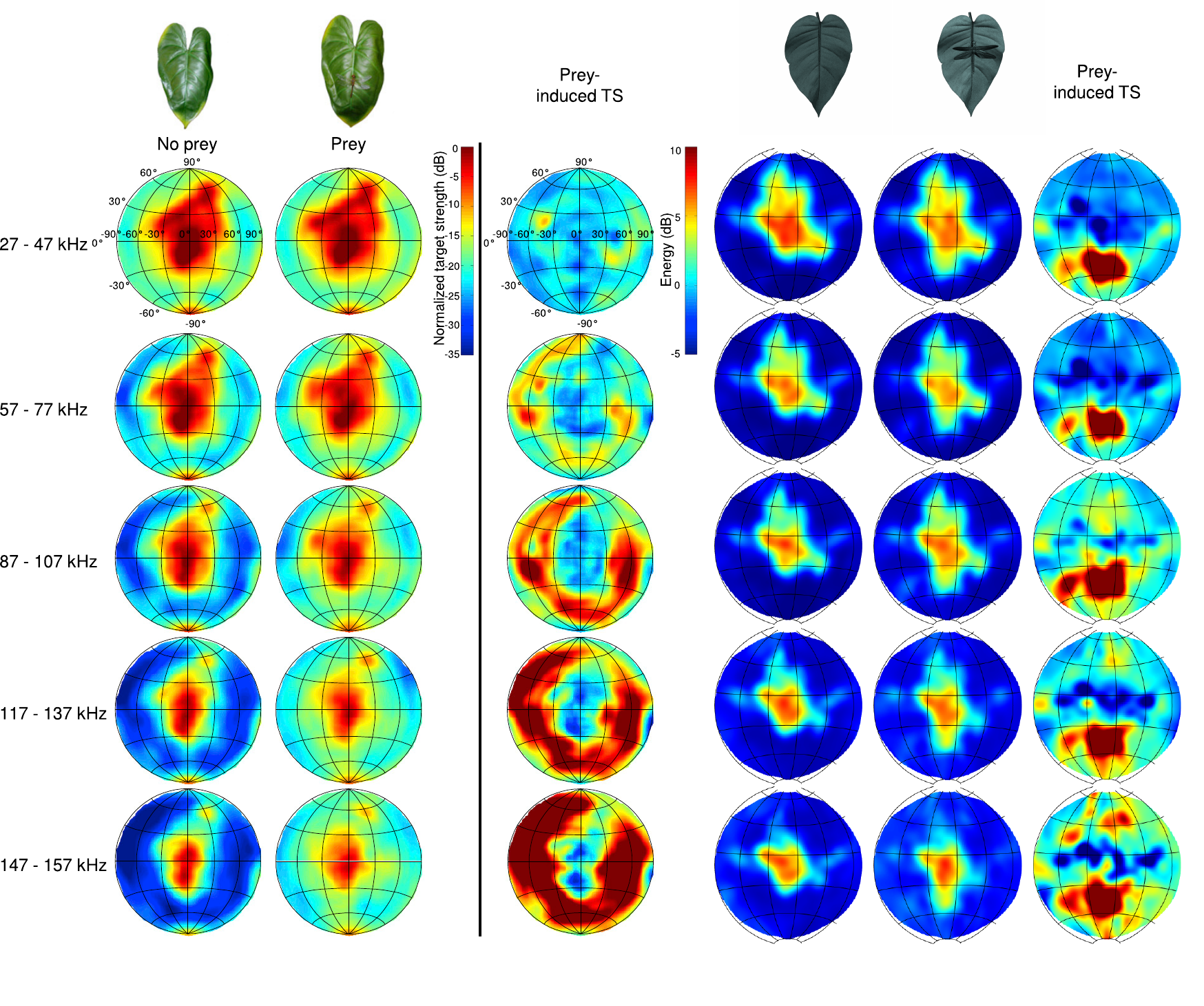}
    \caption{Overview of the validation experiment based on the hunting behavior of Micronycteris \textit{microtus} when hunting silent and motionless prey. The measurements in the left column were taken from \cite{moto:c:irua:161459_geip_bats}, which were performed to measure the spatial reflectivity pattern of leaves with and without prey. The difference between the full and empty leaf is most prominent from oblique angles (indicated by the "Prey-induced Target Strength" column). A simulation setup with similar characteristics (i.e., leaf with or without a dragonfly) was analyzed using SonoTraceLab, which results in reflectivity patterns with similar cues (i.e., larger differences from oblique angles). As these cues are biologically relevant for explaining the hunting behavior of M. \textit{microtus}, this indicates the validity of SonoTraceLab for analyzing biological echolocation scenarios.} 
    \label{fig:mmicrSetup}
\end{figure*}

As a final validation example, we have simulated the spatial reflectivity of a setup relevant to the hunting behavior of Micronycteris \textit{microtus}. This bat hunts in the neotropical forests in Panama and regularly attacks silent and motionless prey \citep{geipel2013perception, Geipel2019, santana2011all}. In previous work, we have discovered the underlying mechanism that supports this behavior, namely the exploitation of the acoustic mirror effect \citep{Geipel2019}, which supports easy prey presence discrimination. In that paper, we performed measurements of the reflectivity pattern of a leaf with and without a dragonfly. We found that the strongest difference between the reflectivity patterns can be found when approaching from oblique angles. These reflectivity patterns can be found in figure \ref{fig:mmicrSetup}. Behavioral experiments showed that the bat is most likely to approach from these oblique angles, maximizing the information obtained from a few measurements. We recreated this setup in SonoTraceLab with a generic model of a leaf \citep{leaf3dmodel} and dragonfly \citep{dragonfly3dmodel} and performed similar reflectivity pattern measurements as in our previous work. From the resulting measurements, we observe similar patterns: a strong reflection when echolocating straight into the leaf and weaker reflections from oblique angles. Furthermore, the difference between the empty and full leaf is also the largest from oblique angles. It should be noted that the simulation is not an exact recreation of the measurements because the leaf shape and dragonfly shape and position are not matched exactly. Therefore, the qualitative comparison is applicable here, as they show that SonoTraceLab can recreate these biologically relevant cues based on 3D models.

\section{Open-source usage}
In this section, we want to detail the open-source aspect of SonoTraceLab. We made the simulation package available through GitHub \citep{githubHttpsGithubCom2024}. The source code repository can be found on Github under \textbf{\url{https://cosyslab.app.link/sonotracelab}} so researchers can further use and validate it in bioacoustics and 3D sonar sensor design. The source code comes with a few example scenarios similar to the experiments performed in this paper, which should allow the users of SonoTraceLab to be up and running efficiently.
\\\\
In terms of usage, the approach is always the same. As a first step, SonoTraceLab loads and prepares the model mesh. This is done by providing a Matlab struct object, which contains the path to the mesh and some configuration parameters, such as its surface material reflection properties and the mesh pose, to the \code{prepareMeshSurface} function. Similarly, a Matlab struct must be made to describe the sensor and the emitter and receiver(s) coordinates. Lastly, a struct must be set up which defines other simulation parameters such as, for example, the directional-sampling density and amount of samples in the impulse response, as well as the speed of sound to be used in the simulation. These three struct objects are the arguments for the function \code{calculateImpulseResponseFast} to perform the specular and diffraction simulation steps. The output of this function will be a Matlab struct containing the impulse responses of all the receivers and for specular and diffraction separately, as well as information about the mesh's vertices that were responsible for the reflections and information such as individual reflection strengths.

\section{Conclusion}
In this paper, we presented SonoTraceLab, a simulation engine written in Matlab that allows the simulation of both technical as well as biological sonar sensor systems using straightforward object meshes. In order to solve the Helmholtz equation, we apply two sub-methods: raytracing for the specular reflections and a Monte Carlo approximation using importance sampling over mesh curvature for the diffraction echos. We provided details into the operation of SonoTraceLab, and explained how the simulator operates. Based on this discussion, we have shown various validation experiments showing the capability of SonoTraceLab to simulate the relevant biological cues present in real-world echolocation scenarios. Finally, we showed how SonoTraceLab can be obtained and used by other researchers through our open-source publication of SonoTraceLab.
\\\\
As stated before, the simulations done with SonoTraceLab can recreate the biologically relevant cues that underpin several echolocation mechanisms found in bats. We believe that by using simulation engines like the one presented here, research in biological echolocation can be significantly accelerated, as many more experiments can be performed in a much shorter time-frame as compared to real-world measurements. Therefore, SonoTraceLab can serve as a hypothesis generation tool, which subsequently can be validated using real-world measurements and biological observations.
\\\\
Despite its applicability, several limitations remain in SonoTraceLab. Firstly, it should be noted that we only approximate the solution of the Helmholtz equation. Therefore, our simulation approach will never replace a real partial differential equation solver. However, as FEM models become computationally intractable quickly, using a system like SonoTraceLab is a valid tool for rapid iterations over various hypotheses. Secondly, SonoTraceLab deals poorly with time-varying mesh geometry, such as fluttering insects. While performing mesh deformations and recalculating the curvature is possible, it is time-consuming. Optimizations could be made in this domain and will be part of the future developments of SonoTraceLab. Furthermore, several more technical issues hinder actual large-scale raytracing simulations (with tens of millions of rays). Migration to other raytracing libraries, such as the NVIDIA OptiX library, would be an interesting step forward. This is also the subject of further developments on SonoTraceLab.

\footnotesize
\section*{Acknowledgements}
The authors would like to thank the Bijzonder OnderzoeksFonds (BOF) of the University of Antwerp for funding this research project. 

\normalsize
\bibliography{references_Zotero,additional_refs}

\begin{thebibliography}{64}
\providecommand{\natexlab}[1]{#1}
\providecommand{\url}[1]{\texttt{#1}}
\expandafter\ifx\csname urlstyle\endcsname\relax
  \providecommand{\doi}[1]{doi: #1}\else
  \providecommand{\doi}{doi: \begingroup \urlstyle{rm}\Url}\fi

\bibitem[Schnitzler et~al.(2003)Schnitzler, Moss, and Denzinger]{schnitzler2003spatial}
Hans-Ulrich Schnitzler, Cynthia~F Moss, and Annette Denzinger.
\newblock From spatial orientation to food acquisition in echolocating bats.
\newblock \emph{Trends in Ecology \& Evolution}, 18\penalty0 (8):\penalty0 386--394, 2003.

\bibitem[Geipel et~al.(2019{\natexlab{a}})Geipel, Steckel, Tschapka, Vanderelst, Schnitzler, Kalko, Peremans, and Simon]{Geipel2019}
I.~Geipel, J.~Steckel, M.~Tschapka, D.~Vanderelst, H.-U. Schnitzler, E.K.V. Kalko, H.~Peremans, and R.~Simon.
\newblock Bats actively use leaves as specular reflectors to detect acoustically camouflaged prey.
\newblock \emph{Current Biology}, 29\penalty0 (16), 2019{\natexlab{a}}.
\newblock ISSN 09609822.
\newblock \doi{10.1016/j.cub.2019.06.076}.

\bibitem[Prat and Yovel(2020)]{prat2020decision}
Yosef Prat and Yossi Yovel.
\newblock Decision making in foraging bats.
\newblock \emph{Current Opinion in Neurobiology}, 60:\penalty0 169--175, 2020.

\bibitem[Stidsholt et~al.(2021)Stidsholt, Greif, Goerlitz, Beedholm, Macaulay, Johnson, and Madsen]{stidsholt2021hunting}
Laura Stidsholt, Stefan Greif, Holger~R Goerlitz, Kristian Beedholm, Jamie Macaulay, Mark Johnson, and Peter~Teglberg Madsen.
\newblock Hunting bats adjust their echolocation to receive weak prey echoes for clutter reduction.
\newblock \emph{Science Advances}, 7\penalty0 (10):\penalty0 eabf1367, 2021.

\bibitem[McGowan and Kloepper(2020)]{mcgowan2020different}
Kathryn~A McGowan and Laura~N Kloepper.
\newblock Different as night and day: Wild bats modify echolocation in complex environments when visual cues are present.
\newblock \emph{Animal Behaviour}, 168:\penalty0 1--6, 2020.

\bibitem[Simon et~al.(2006)Simon, Holderied, and Von~Helversen]{simon2006size}
Ralph Simon, Marc~W Holderied, and Otto Von~Helversen.
\newblock Size discrimination of hollow hemispheres by echolocation in a nectar feeding bat.
\newblock \emph{Journal of experimental biology}, 209\penalty0 (18):\penalty0 3599--3609, 2006.

\bibitem[Simon et~al.(2011)Simon, Holderied, Koch, and {von Helversen}]{simon2011floral}
Ralph Simon, Marc~W Holderied, Corinna~U Koch, and Otto {von Helversen}.
\newblock Floral acoustics: Conspicuous echoes of a dish-shaped leaf attract bat pollinators.
\newblock \emph{Science (New York, N.Y.)}, 333\penalty0 (6042):\penalty0 631--633, 2011.

\bibitem[Simon et~al.(2021{\natexlab{a}})Simon, Bakunowski, {Reyes-Vasques}, Tschapka, Kn{\"o}rnschild, Steckel, and Stowell]{simon2021acoustic}
Ralph Simon, Karol Bakunowski, Angel~Eduardo {Reyes-Vasques}, Marco Tschapka, Mirjam Kn{\"o}rnschild, Jan Steckel, and Dan Stowell.
\newblock Acoustic traits of bat-pollinated flowers compared to flowers of other pollination syndromes and their echo-based classification using convolutional neural networks.
\newblock \emph{PLoS Computational Biology}, 17\penalty0 (12):\penalty0 e1009706, 2021{\natexlab{a}}.

\bibitem[Sch{\"o}ner et~al.(2015{\natexlab{a}})Sch{\"o}ner, Sch{\"o}ner, Simon, Grafe, Puechmaille, Ji, and Kerth]{schoner2015bats}
Michael~G Sch{\"o}ner, Caroline~R Sch{\"o}ner, Ralph Simon, T~Ulmar Grafe, S{\'e}bastien~J Puechmaille, Liaw~Lin Ji, and Gerald Kerth.
\newblock Bats are acoustically attracted to mutualistic carnivorous plants.
\newblock \emph{Current Biology}, 25\penalty0 (14):\penalty0 1911--1916, 2015{\natexlab{a}}.

\bibitem[Geipel et~al.(2020)Geipel, Kernan, Litterer, Carter, Page, and Ter~Hofstede]{geipel2020predation}
Inga Geipel, Ciara~E Kernan, Amber~S Litterer, Gerald~G Carter, Rachel~A Page, and Hannah~M Ter~Hofstede.
\newblock Predation risks of signalling and searching: Bats prefer moving katydids.
\newblock \emph{Biology Letters}, 16\penalty0 (4):\penalty0 20190837, 2020.

\bibitem[Geipel et~al.(2013)Geipel, Jung, and Kalko]{geipel2013perception}
Inga Geipel, Kirsten Jung, and Elisabeth~KV Kalko.
\newblock Perception of silent and motionless prey on vegetation by echolocation in the gleaning bat {{Micronycteris}} microtis.
\newblock \emph{Proceedings of the Royal Society B: Biological Sciences}, 280\penalty0 (1754):\penalty0 20122830, 2013.

\bibitem[Stilz(2017)]{stilz2017glass}
Peter Stilz.
\newblock How glass fronts deceive bats.
\newblock \emph{Science (New York, N.Y.)}, 357\penalty0 (6355):\penalty0 977--978, 2017.

\bibitem[Greif and Siemers(2010)]{greif2010innate}
Stefan Greif and Bj{\"o}rn~M Siemers.
\newblock Innate recognition of water bodies in echolocating bats.
\newblock \emph{Nature communications}, 1\penalty0 (1):\penalty0 107, 2010.

\bibitem[Sch{\"o}ner et~al.(2015{\natexlab{b}})Sch{\"o}ner, Sch{\"o}ner, Simon, Grafe, Puechmaille, Ji, and Kerth]{schonerBatsAreAcoustically2015}
Michael~G. Sch{\"o}ner, Caroline~R. Sch{\"o}ner, Ralph Simon, T.~Ulmar Grafe, S{\'e}bastien~J. Puechmaille, Liaw~Lin Ji, and Gerald Kerth.
\newblock Bats are acoustically attracted to mutualistic carnivorous plants.
\newblock \emph{Current Biology}, 25\penalty0 (14):\penalty0 1911--1916, 2015{\natexlab{b}}.

\bibitem[Geipel et~al.(2019{\natexlab{b}})Geipel, Steckel, Tschapka, Vanderelst, Schnitzler, Kalko, Peremans, and Simon]{moto:c:irua:161459_geip_bats}
Inga Geipel, Jan Steckel, Marco Tschapka, Dieter Vanderelst, Hans-Ulrich Schnitzler, Elisabeth~K.V. Kalko, Herbert Peremans, and Ralph Simon.
\newblock Bats actively use leaves as specular reflectors to detect acoustically camouflaged prey.
\newblock \emph{Current biology}, 29\penalty0 (16):\penalty0 2731--2736, 2019{\natexlab{b}}.
\newblock ISSN 1879-0445.
\newblock \doi{10.1016/J.CUB.2019.06.076}.

\bibitem[{\"U}bernickel et~al.(2016){\"U}bernickel, Simon, Kalko, and Tschapka]{ubernickelSensoryChallengesTrawling2016}
Kirstin {\"U}bernickel, Ralph Simon, Elisabeth~KV Kalko, and Marco Tschapka.
\newblock Sensory challenges for trawling bats: {{Finding}} transient prey on water surfaces.
\newblock \emph{The Journal of the Acoustical Society of America}, 139\penalty0 (4):\penalty0 1914--1922, 2016.

\bibitem[Fontaine and Peremans(2011)]{fontaineCompressiveSensingStrategy2011}
Bertrand Fontaine and Herbert Peremans.
\newblock Compressive sensing: {{A}} strategy for fluttering target discrimination employed by bats emitting broadband calls.
\newblock \emph{The Journal of the Acoustical Society of America}, 129\penalty0 (2):\penalty0 1100--1110, 2011.

\bibitem[Moss and Zagaeski(1994)]{mossAcousticInformationAvailable1994}
Cynthia~F. Moss and Mark Zagaeski.
\newblock Acoustic information available to bats using frequency-modulated sounds for the perception of insect prey.
\newblock \emph{The Journal of the Acoustical Society of America}, 95\penalty0 (5):\penalty0 2745--2756, 1994.

\bibitem[Caspers and M{\"u}ller(2018)]{caspersDesignDynamicBiomimetic2018}
Philip Caspers and Rolf M{\"u}ller.
\newblock A design for a dynamic biomimetic sonarhead inspired by horseshoe bats.
\newblock \emph{Bioinspiration \& biomimetics}, 13\penalty0 (4):\penalty0 046011, 2018.

\bibitem[Schillebeeckx et~al.(2011)Schillebeeckx, De~Mey, Vanderelst, and Peremans]{schillebeeckxBiomimeticSonarBinaural2011a}
Filips Schillebeeckx, Fons De~Mey, Dieter Vanderelst, and Herbert Peremans.
\newblock Biomimetic sonar: {{Binaural 3D}} localization using artificial bat pinnae.
\newblock \emph{The International Journal of Robotics Research}, 30\penalty0 (8):\penalty0 975--987, 2011.

\bibitem[Sutlive et~al.(2020)Sutlive, Singh, Zhang, and M{\"u}ller]{sutliveBiomimeticSoftRobotic2020}
Joseph Sutlive, Agoshpreet Singh, Shuxin Zhang, and Rolf M{\"u}ller.
\newblock A biomimetic soft robotic pinna for emulating dynamic reception behavior of horseshoe bats.
\newblock \emph{Bioinspiration \& Biomimetics}, 16\penalty0 (1):\penalty0 016016, 2020.

\bibitem[Steckel and Peremans(2012)]{moto:c:irua:100095_stec_nove}
Jan Steckel and Herbert Peremans.
\newblock A novel biomimetic sonarhead using beamforming technology to mimic bat echolocation.
\newblock \emph{IEEE transactions on ultrasonics, ferroelectrics and frequency control}, 59\penalty0 (7):\penalty0 1369--1377, 2012.
\newblock ISSN 0885-3010.
\newblock \doi{10.1109/TUFFC.2012.2337}.

\bibitem[Steckel et~al.(2011)Steckel, Schillebeeckx, and Peremans]{moto:c:irua:92037_stec_biom}
Jan Steckel, Filips Schillebeeckx, and Herbert Peremans.
\newblock Biomimetic sonar, outer ears versus arrays.
\newblock In \emph{Proceedings {{IEEE SENSORS}} 2011 Conference, October 2011, Limerick, Ireland}, pages 821--824. 2011.
\newblock ISBN 978-1-4244-9288-6.

\bibitem[Siemers and Schnitzler(2004)]{siemersEcholocationSignalsReflect2004}
Bj{\"o}rn~M. Siemers and Hans-Ulrich Schnitzler.
\newblock Echolocation signals reflect niche differentiation in five sympatric congeneric bat species.
\newblock \emph{Nature}, 429\penalty0 (6992):\penalty0 657--661, 2004.

\bibitem[Simon et~al.(2021{\natexlab{b}})Simon, Bakunowski, {Reyes-Vasques}, Tschapka, Knoernschild, Steckel, and Stowell]{moto:c:irua:184867_simo_acou}
Ralph Simon, Karol Bakunowski, Angel~Eduardo {Reyes-Vasques}, Marco Tschapka, Mirjam Knoernschild, Jan Steckel, and Dan Stowell.
\newblock Acoustic traits of bat-pollinated flowers compared to flowers of other pollination syndromes and their echo-based classification using convolutional neural networks.
\newblock \emph{PLoS computational biology}, 17\penalty0 (12):\penalty0 20 p., 2021{\natexlab{b}}.
\newblock ISSN 1553-7358.
\newblock \doi{10.1371/JOURNAL.PCBI.1009706}.

\bibitem[Nguyen et~al.(2021)Nguyen, Vanderelst, and Peremans]{nguyenSensorimotorBehaviorInformational2021}
Thinh Nguyen, Dieter Vanderelst, and Herbert Peremans.
\newblock Sensorimotor behavior under informational constraints: A robotic model of prey localization in the bat {{Micronycteris}} microtis.
\newblock In \emph{Artificial {{Life Conference Proceedings}} 33}, volume 2021, page~61. MIT Press One Rogers Street, Cambridge, MA 02142-1209, USA journals-info~{\dots}, 2021.
\newblock ISBN 2693-1508.

\bibitem[Arimoto et~al.(2005)Arimoto, Sekimoto, Hashiguchi, and Ozawa]{arimotoNaturalResolutionIllposedness2005}
Suguru Arimoto, Masahiro Sekimoto, Hiroe Hashiguchi, and Ryuta Ozawa.
\newblock Natural resolution of ill-posedness of inverse kinematics for redundant robots: {{A}} challenge to {{Bernstein}}'s degrees-of-freedom problem.
\newblock \emph{Advanced Robotics}, 19\penalty0 (4):\penalty0 401--434, 2005.

\bibitem[{Cosys-Lab}(2018)]{cosys-labMicronycterisMicrotisCapturing2018}
{Cosys-Lab}.
\newblock Micronycteris {{Microtis}} capturing a dragonfly, November 2018.

\bibitem[Pierce(2019)]{pierceAcousticsIntroductionIts2019}
Allan~D. Pierce.
\newblock \emph{Acoustics: An Introduction to Its Physical Principles and Applications}.
\newblock Springer, 2019.
\newblock ISBN 3-030-11214-4.

\bibitem[Everstine(1997)]{everstineFiniteElementFormulatons1997}
G.~C. Everstine.
\newblock Finite element formulatons of structural acoustics problems.
\newblock \emph{Computers \& Structures}, 65\penalty0 (3):\penalty0 307--321, 1997.

\bibitem[Ihlenburg(1998)]{ihlenburgFiniteElementAnalysis1998}
Frank Ihlenburg.
\newblock \emph{Finite Element Analysis of Acoustic Scattering}.
\newblock Springer, 1998.

\bibitem[Thompson(2006)]{thompsonReviewFiniteelementMethods2006}
Lonny~L. Thompson.
\newblock A review of finite-element methods for time-harmonic acoustics.
\newblock \emph{The Journal of the Acoustical Society of America}, 119\penalty0 (3):\penalty0 1315--1330, 2006.

\bibitem[Kirkup(2007)]{kirkupBoundaryElementMethod2007}
Stephen Kirkup.
\newblock \emph{The Boundary Element Method in Acoustics}.
\newblock -, 2007.

\bibitem[Kirkup(2019)]{kirkupBoundaryElementMethod2019}
Stephen Kirkup.
\newblock The boundary element method in acoustics: {{A}} survey.
\newblock \emph{Applied Sciences}, 9\penalty0 (8):\penalty0 1642, 2019.

\bibitem[Onaka et~al.(2009)Onaka, Morise, and Nishiura]{onakaDesign3dimensionalSound2009}
Kenji Onaka, Masanori Morise, and Takanobu Nishiura.
\newblock A design of 3-dimensional sound field simulator based on acoustic ray tracing and {{HRTF}}.
\newblock In \emph{2009 {{IEEE}} 13th {{International Symposium}} on {{Consumer Electronics}}}, pages 233--236. IEEE, 2009.
\newblock ISBN 1-4244-2975-7.

\bibitem[R{\"o}ber et~al.(2006)R{\"o}ber, Andres, and Masuch]{roberHRTFSimulationsAcoustic2006}
Niklas R{\"o}ber, Sven Andres, and Maic Masuch.
\newblock \emph{{{HRTF}} Simulations through Acoustic Raytracing}.
\newblock Universit{\"a}ts-und Landesbibliothek Sachsen-Anhalt, 2006.

\bibitem[R{\"o}ber et~al.(2007)R{\"o}ber, Kaminski, and Masuch]{roberRayAcousticsUsing2007}
Niklas R{\"o}ber, Ulrich Kaminski, and Maic Masuch.
\newblock Ray acoustics using computer graphics technology.
\newblock In \emph{10th {{International Conference}} on {{Digital Audio Effects}} ({{DAFx-07}}), {{S}}}, pages 117--124, 2007.

\bibitem[COMSOL()]{comsolcomsolCOMSOLMultiphysicsSoftware}
COMSOL COMSOL.
\newblock {{COMSOL}}: {{Multiphysics Software}} for {{Optimizing Designs}}.
\newblock https://www.comsol.com/.

\bibitem[Siemens()]{siemensSiemensNXAcoustics}
{\relax NX}~Siemens.
\newblock Siemens {{NX Acoustics}}.
\newblock https://plm.sw.siemens.com/en-US/simcenter/simulation-test/acoustic-simulation/.

\bibitem[Treeby and Cox(2010)]{treebyKWaveMATLABToolbox2010}
Bradley~E. Treeby and Benjamin~T. Cox.
\newblock K-{{Wave}}: {{MATLAB}} toolbox for the simulation and reconstruction of photoacoustic wave fields.
\newblock \emph{Journal of biomedical optics}, 15\penalty0 (2):\penalty0 021314--021314--12, 2010.

\bibitem[{Field-II}()]{field-iiFieldIIUltrasound}
FIELD-II {Field-II}.
\newblock Field {{II Ultrasound Simulation Program}}.
\newblock https://field-ii.dk/.

\bibitem[Adobe()]{adobeSTLFilesExplained}
{\relax STL}~Adobe.
\newblock {{STL}} files explained {\textbar} {{Learn}} about the {{STL}} file format {\textbar} {{Adobe}}.
\newblock https://www.adobe.com/au/creativecloud/file-types/image/vector/stl-file.html.

\bibitem[Matlab()]{matlabRemoveSurfaceMesh}
Lidar~Toolbox Matlab.
\newblock Remove surface mesh defects - {{MATLAB removeDefects}} - {{MathWorks Benelux}}.

\bibitem[Rusinkiewicz(2004)]{rusinkiewiczEstimatingCurvaturesTheir2004}
Szymon Rusinkiewicz.
\newblock Estimating curvatures and their derivatives on triangle meshes.
\newblock In \emph{Proceedings. 2nd {{International Symposium}} on {{3D Data Processing}}, {{Visualization}} and {{Transmission}}, 2004. {{3DPVT}} 2004.}, pages 486--493. IEEE, 2004.
\newblock ISBN 0-7695-2223-8.

\bibitem[Binek et~al.(2019)Binek, Pilch, and Kamisinski]{binekApplicationFittedMeasurements2019}
Wojciech Binek, Adam Pilch, and Tadeusz Kamisinski.
\newblock Application of fitted to measurements sound reflecion model in geometrical acoustics simulations.
\newblock In \emph{{{INTER-NOISE}} and {{NOISE-CON Congress}} and {{Conference Proceedings}}}, volume 259, pages 1189--1194. Institute of Noise Control Engineering, 2019.
\newblock ISBN 0736-2935.

\bibitem[Durany et~al.(2015)Durany, Mateos, and Garriga]{duranyAnalyticalComputationAcoustic2015}
Jaume Durany, Toni Mateos, and Adan Garriga.
\newblock Analytical computation of acoustic bidirectional reflectance distribution functions.
\newblock \emph{Open Journal of Acoustics}, 5\penalty0 (04):\penalty0 207, 2015.

\bibitem[Matusik et~al.(2003)Matusik, Pfister, Brand, and McMillan]{matusikEfficientIsotropicBRDF2003}
Wojciech Matusik, Hanspeter Pfister, Matthew Brand, and Leonard McMillan.
\newblock Efficient isotropic {{BRDF}} measurement.
\newblock 2003.

\bibitem[Kerstens et~al.(2019{\natexlab{a}})Kerstens, Laurijssen, and Steckel]{moto:c:irua:165188_kers_a}
Robin Kerstens, Dennis Laurijssen, and Jan Steckel.
\newblock {{eRTIS}}: A fully embedded real time {{3D}} imaging sonar sensor for robotic applications.
\newblock In \emph{{{IEEE}} International Conference on Robotics and Automation}, pages 1438--1443. Ieee, 2019{\natexlab{a}}.
\newblock ISBN 978-1-5386-6026-3.
\newblock \doi{10.1109/ICRA.2019.8794419}.

\bibitem[Kerstens et~al.(2019{\natexlab{b}})Kerstens, Laurijssen, Schouten, and Steckel]{moto:c:irua:166446_kers_3d}
Robin Kerstens, Dennis Laurijssen, Girmi Schouten, and Jan Steckel.
\newblock {{3D}} point cloud data acquisition using a synchronized in-air imaging sonar sensor network.
\newblock In \emph{Proceedings of the {{International Conference}} on {{Intelligent Robots}} and {{Systems}}}, pages 5855--5861. IEEE, 2019{\natexlab{b}}.
\newblock ISBN 978-1-72814-004-9.
\newblock \doi{10.1109/IROS40897.2019.8968469}.

\bibitem[Verellen et~al.(2020)Verellen, Kerstens, Laurijssen, and Steckel]{moto:c:irua:168766_vere_urti}
Thomas Verellen, Robin Kerstens, Dennis Laurijssen, and Jan Steckel.
\newblock {{URTIS}} : A small {{3D}} imaging sonar sensor for robotic applications.
\newblock In \emph{Proceedings of the ... {{IEEE International Conference}} on {{Acoustics}}, {{Speech}}, and {{Signal Processing}}}, pages 4801--4805. IEEE, 2020.
\newblock ISBN 978-1-5090-6631-5.
\newblock \doi{10.1109/ICASSP40776.2020.9053536}.

\bibitem[Balemans et~al.(2021)Balemans, Hellinckx, Latr{\'e}, Reiter, and Steckel]{moto:c:irua:186389_bale_sens}
Niels Balemans, Peter Hellinckx, Steven Latr{\'e}, Philippe Reiter, and Jan Steckel.
\newblock {{S2L-SLAM}} : Sensor fusion driven {{SLAM}} using sonar, {{LiDAR}} and deep neural networks.
\newblock In \emph{Proceedings of {{IEEE Sensors}}}, page 4 p. Ieee, 2021.
\newblock ISBN 978-1-72819-501-8.
\newblock \doi{10.1109/SENSORS47087.2021.9639772}.

\bibitem[Thomas et~al.(2004)Thomas, Moss, and Vater]{thomasEcholocationBatsDolphins2004}
Jeanette~A. Thomas, Cynthia~F. Moss, and Marianne Vater.
\newblock \emph{Echolocation in Bats and Dolphins}.
\newblock University of Chicago press, 2004.
\newblock ISBN 0-226-79599-3.

\bibitem[Shumskiy(2013)]{shumskiyGpuRayTracing2013}
Vladimir Shumskiy.
\newblock Gpu ray tracing--comparative study on ray-triangle intersection algorithms.
\newblock In \emph{Transactions on {{Computational Science XIX}}: {{Special Issue}} on {{Computer Graphics}}}, pages 78--91. Springer, 2013.
\newblock ISBN 3-642-39758-1.

\bibitem[Corsini et~al.(2012)Corsini, Cignoni, and Scopigno]{corsiniEfficientFlexibleSampling2012}
Massimiliano Corsini, Paolo Cignoni, and Roberto Scopigno.
\newblock Efficient and flexible sampling with blue noise properties of triangular meshes.
\newblock \emph{IEEE transactions on visualization and computer graphics}, 18\penalty0 (6):\penalty0 914--924, 2012.

\bibitem[De~Mey et~al.(2010)De~Mey, Schillebeeckx, Vanderelst, Boen, and Peremans]{demeyModellingSimultaneousEcho2010}
Fons De~Mey, Filips Schillebeeckx, Dieter Vanderelst, Andre Boen, and Herbert Peremans.
\newblock Modelling simultaneous echo waveform reconstruction and localization in bats.
\newblock \emph{Biosystems}, 100\penalty0 (2):\penalty0 94--100, 2010.

\bibitem[Obrist et~al.(1993)Obrist, Fenton, Eger, and Schlegel]{obristWhatEarsBats1993}
Martin~K. Obrist, M.~Brock Fenton, Judith~L. Eger, and Peter~A. Schlegel.
\newblock What ears do for bats: A comparative study of pinna sound pressure transformation in {{Chiroptera}}.
\newblock \emph{Journal of Experimental Biology}, 180\penalty0 (1):\penalty0 119--152, 1993.

\bibitem[Vanderelst et~al.(2010)Vanderelst, De~Mey, Peremans, Geipel, Kalko, and Firzlaff]{vanderelstWhatNoseleavesFM2010}
Dieter Vanderelst, Fons De~Mey, Herbert Peremans, Inga Geipel, Elisabeth Kalko, and Uwe Firzlaff.
\newblock What noseleaves do for {{FM}} bats depends on their degree of sensorial specialization.
\newblock \emph{PloS one}, 5\penalty0 (8):\penalty0 e11893, 2010.

\bibitem[Leopardi(2006)]{leopardiPartitionUnitSphere2006}
Paul Leopardi.
\newblock A partition of the unit sphere into regions of equal area and small diameter.
\newblock \emph{Electronic Transactions on Numerical Analysis}, 25\penalty0 (12):\penalty0 309--327, 2006.

\bibitem[De~Mey et~al.(2008)De~Mey, Reijniers, Peremans, Otani, and Firzlaff]{demeySimulatedHeadRelated2008}
F.~De~Mey, J.~Reijniers, H.~Peremans, M.~Otani, and U.~Firzlaff.
\newblock Simulated head related transfer function of the phyllostomid bat {{Phyllostomus}} discolor.
\newblock \emph{The Journal of the Acoustical Society of America}, 124\penalty0 (4):\penalty0 2123--2132, 2008.

\bibitem[Simon et~al.(2020)Simon, Rupitsch, Baumann, Wu, Peremans, and Steckel]{moto:c:irua:165435_simo_bioi}
Ralph Simon, Stefan Rupitsch, Markus Baumann, Huan Wu, Herbert Peremans, and Jan Steckel.
\newblock Bioinspired sonar reflectors as guiding beacons for autonomous navigation.
\newblock \emph{Proceedings of the National Academy of Sciences of the United States of America}, 117\penalty0 (3):\penalty0 1367--1374, 2020.
\newblock ISSN 0027-8424.
\newblock \doi{10.1073/PNAS.1909890117}.

\bibitem[Santana et~al.(2011)Santana, Geipel, Dumont, Kalka, and Kalko]{santana2011all}
Sharlene~E Santana, Inga Geipel, Elizabeth~R Dumont, Margareta~B Kalka, and Elisabeth~KV Kalko.
\newblock All you can eat: High performance capacity and plasticity in the common big-eared bat, {{Micronycteris}} microtis ({{Chiroptera}}: {{Phyllostomidae}}).
\newblock \emph{PLoS One}, 6\penalty0 (12):\penalty0 e28584, 2011.

\bibitem[{Diana Liu}(2019)]{leaf3dmodel}
{Diana Liu}.
\newblock Leaf 3d model.
\newblock This work is licensed under the Creative Commons Attribution 4.0 International License., 2019.
\newblock URL \url{https://skfb.ly/6SuSN}.

\bibitem[{ffish.asia}(2023)]{dragonfly3dmodel}
{ffish.asia}.
\newblock Anax parthenope 3d model.
\newblock This work is licensed under the Creative Commons Attribution-NonCommercial-NoDerivs 4.0 International License., 2023.
\newblock URL \url{https://skfb.ly/oJNSP}.

\bibitem[Github(2024)]{githubHttpsGithubCom2024}
Cosys-lab Github.
\newblock {{https://github.com/Cosys-Lab/SonoTraceLab}}.
\newblock Cosys-Lab, March 2024.

\end{thebibliography}

\end{document}